\documentclass[osajnl,twocolumn,showpacs,superscriptaddress,10pt]{revtex4-1} %% use 11pt for Applied Optics
\usepackage{amsmath,amssymb,graphicx}
\begin{document}

\title{Generation and tomography of arbitrary qubit states in a transient collective atomic excitation}

\author{T.~Brannan}
\affiliation{Institute for Quantum Science and Technology, University of Calgary, Calgary, Alberta T2N 1N4, Canada}

\author{Z.~Qin}
\affiliation{Institute for Quantum Science and Technology, University of Calgary, Calgary, Alberta T2N 1N4, Canada}
\affiliation{Quantum Institute for Light and Atoms, State Key Laboratory of Precision Spectroscopy, East China Normal University, Shanghai 200062, People's Republic of China}

\author{A.~MacRae}
\affiliation{Institute for Quantum Science and Technology, University of Calgary, Calgary, Alberta T2N 1N4, Canada}
\affiliation{Department of Physics, University of California, Berkeley, CA 94720}

\author{A.~I.~Lvovsky}\email{Corresponding author: LVOV@ucalgary.ca}
\affiliation{Institute for Quantum Science and Technology, University of Calgary, Calgary, Alberta T2N 1N4, Canada}
\affiliation{Russian Quantum Center, 100 Novaya St., Skolkovo, Moscow 143025, Russia}

\begin{abstract}
	We demonstrate the generation of heralded Fock-basis qubits $(a|0\rangle + b|1\rangle)$ from transient collective spin excitations in a hot atomic vapor. The collective spin excitation is heralded by Raman-scattered photons in a four-wave mixing process seeded by a weak coherent optical excitation. The amplitude and phase of the seed field allow arbitrary control over the qubit coefficients. The state of the ensemble is read out optically and then characterized using balanced homodyne tomography. This work represents a step towards quantum engineering of collective atomic spin states.
\end{abstract}

%\ocis{(270.0270) Quantum optics; (270.5585) Quantum information and processing; (270.4180) Multiphoton processes; (190.4380) Nonlinear optics, four-wave mixing.}% REPLACE WITH CORRECT OCIS CODES FOR YOUR ARTICLE
%                          % NOTE: \ocis{} IS ALIASED TO \pacs{} BUT MUST
%                          % FORMAT THE TERMS CORRECTLY FOR EACH JOURNAL

\maketitle %% required

%\section{Introduction}

\noindent Quantum state engineering remains a central challenge for the development of quantum technologies. There has been a great deal of progress in preparing and measuring interesting quantum states in the optical domain, driven by the attractiveness of light as a mediator for quantum information and quantum communication \cite{Lvovsky_review}. Quantum state engineering has also been demonstrated in many other quantum systems \cite{Blatt13}, including superconducting circuits \cite{microwave} and trapped ion ensembles \cite{trapped_ion}.

Extending these techniques to collective spin excitations (CSEs) in atomic ensembles would prove valuable, as atomic systems have shown to be a promising candidate for storage of quantum states and light-matter interations \cite{Hammerer_review}. Additionally, engineering arbitrary quantum states of these atomic CSEs can find applications in quantum memories \cite{lvov_memory}, quantum repeaters \cite{DLCZ}, quantum logic gates \cite{logic_gates}, and quantum metrology \cite{metrology}.

%Until recently, engineering of CSE states has been limited by squeezed states prepared by means of non-demolition quadrature measurements\cite{sq1, sq2, sq3}.
Such engineering is enabled by the recently developed methods of preparing the single-quantum CSE by heralding on a Raman-scattered photon \cite{DLCZ} as well as applying the ideology of homodyne tomography for measurement of CSE states \cite{sq3,MacRae}. Used together \cite{MacRae,Polzik}, these techniques permit extension of well developed strategies of quantum state engineering from the optical domain to atomic CSEs. In this work we demonstrate creation and characterization of an arbitrary qubit state of a transient CSE --- a superposition of the vacuum and single-quantum states. This represents an important step towards full control over the CSE Hilbert space.
%Recent progress with Rydberg atoms has also demonstrated superpositions of zero and one Rydberg excitations \cite{Kuzmich}.

%\section{Generation of CSE qubits}
Our system employs coherent double Raman scattering (four-wave mixing) in an ensemble of three-level $\Lambda$-type atoms (Fig.~1(b)), akin to the DLCZ protocol \cite{DLCZ}. If all atoms are initially in one ground state $|b\rangle$, emission  of a single Raman-scattered photon corresponds to a ``write" event where one atom has transitioned to the other ground state $|c\rangle$. Since the atoms are indistinguishable, the excitation occurs collectively over all of the atoms within the interaction region, leaving the atomic ensemble in the single-quantum CSE state:
\begin{equation}\label{CSE1}
|1\rangle = \frac{1}{\sqrt N} \sum\limits_{n=1}^{N} e^{i\phi_n}  |b_1,\ldots b_{n-1}c_n b_{n+1}\ldots b_N\rangle
\end{equation}
Here, $N$ is the total number of interacting atoms and $\phi_n$ gives the phase associated with the recoil of each atom. The Hamiltonian for the Raman scattering is
\begin{equation}\label{Ham}
\hat{H} = \gamma (\hat{a}_i\hat{a}_{\rm CSE} + \hat{a}^\dagger_i\hat{a}^\dagger_{\rm CSE}),
\end{equation}
where $\hat{a}_i$ and $\hat{a}_{\rm CSE}$ are the annihilation operators for the scattered photon and the atomic CSE respectively, and $\gamma$ is the coupling constant.

This Hamiltonian is identical to the usual two-mode squeezing Hamiltonian characteristic of spontaneous parametric down-conversion, however here only one of the modes is optical while the other one is atomic. The entanglement it creates enables us to engineer quantum states of the CSE by performing measurements on the optical mode. 

%This single-quantum CSE is sufficient for the generation of a CSE qubit, but higher order excitations can also be obtained by heralding on multiple Raman-scattered photons.

In DLCZ, detection of the photon heralds the creation of a single-quantum CSE, possibly delocalized between two ensembles. More sophisticated measurements, however, permit engineering of more complex atomic states. We experimentally generate arbitrary Fock-basis qubits, following the technique reminiscent of Bimbard {\it et al.} \cite{Bimbard}. We seed the optical channel of Hamiltonian (\ref{Ham}) by a weak coherent state $|\alpha\rangle$. Assuming that the initial atomic state is vacuum (all atoms in $|b\rangle$), interaction under Hamiltonian (\ref{Ham}) for time $t$ leads, in the first order in $\alpha$ and $\gamma$, to the following state:
\begin{equation}\label{Psiout}
|\Psi_{\rm out}\rangle=|0_i,0_{\rm CSE}\rangle+\alpha|1_i,0_{\rm CSE}\rangle-i\frac{\gamma t}{\hbar}|1_i,1_{\rm CSE}\rangle.
\end{equation}
In writing Eq.~(\ref{Psiout}), we used $|\alpha\rangle\approx|0\rangle+\alpha|1\rangle$ and $e^{-i\hat H t/\hbar}\approx\hat I-i\hat H t/\hbar$.
Now if we perform photon detection in the optical mode, a single detection event will occur with probability
\begin{equation}\label{prcount}
{\rm pr_{count}}=|\alpha|^2+(|\gamma| t/\hbar)^2
\end{equation}
which projects the CSE mode onto
\begin{equation}\label{psi}
|\Psi_{\rm CSE}\rangle=\alpha|0\rangle-i\frac{\gamma t}{\hbar}|1\rangle.
\end{equation}
In other words, because the photon detector  cannot distinguish between a click coming from the coherent beam or from Raman scattering, the state of the CSE collapases into a superposition of matter states corresponding to the situations where Raman scattering has and has not occurred.  The relative amplitudes of the vacuum and single-photon terms in Eq.~(\ref{psi}) can be controlled by the amplitude and phase of the weak coherent beam, allowing for the creation of any arbitrary single-rail atomic qubit.

Subsequent excitation of the $|c\rangle \rightarrow |a\rangle$ transition permits the CSE state to be ``read out". Atoms return to the original state $|b\rangle$ and the CSE is transformed into an optical state coherently emitted along the $|a\rangle \rightarrow |b\rangle$ transition. Constructive interference for this read transition occurs when phase-matching conditions are met, collectively enhancing the emission into a specific optical mode \cite{DLCZ}. The resulting optical state can be measured using homodyne tomography, thereby allowing the characterization of the CSE.

In our experiment, a single, strong continuous laser pumps both of the read and write transitions involved in DLCZ simultaneously. The read and write transitions, with the associated emission of the photons (which we call ``signal" and ``idler" in analogy to spontaneous parametric down conversion) occur at the same time, so the CSE is of transient character \cite{MacRae}. In future experiments, the CSE lifetime can be extended by switching to pulsed excitations, allowing for time separation of the read and write events.

%\section{Experimental methods}
Our $\Lambda$ system employs the 795 nm D1 multiplet in $^{85}$Rb, shown in Fig.~1(b). Both optical transitions are driven by a single laser, which is blue-detuned by 0.8 GHz from the $|5S_{1/2},F=2\rangle\to |5P_{1/2}\rangle$ ($|b\rangle\to|a\rangle$) transition and by 3.9 GHz from the $|5S_{1/2},F=3\rangle\to |5P_{1/2}\rangle$ ($|c\rangle\to|a\rangle$) transition. This detuning is chosen to avoid absorption losses while retaining reasonable non-linearity. %Excitation of this   $|5S_{1/2},F=3\rangle$  ($|b\rangle$) state.
%Measurement of a Raman (``idler") photon then heralds the excitation of one of the atoms into the  $|c\rangle$ state. This CSE then collectively enhances the Raman transition back to $|b\rangle$ , immediately completing optical readout of the atomic state into the ``signal" mode.
%Optical pumping drives most of the atoms into the

%\begin{figure}[tb]
%\centering
%\includegraphics[width = 0.40\columnwidth]{Fig1-Lambda_rb2.pdf}
%\caption{The three-level lambda system in $^{85}$Rb. A strong pump drives both Raman transitions between the $5S_{1/2},F=2$ and $F=3$ states, resulting in emission of photon pairs into the signal and idler channels. %The probability of this second Raman transition returning the atoms to the original state is collectively enhanced along the phase-matched direction, leading to high-efficiency readout of the signal photon.
%}
%\end{figure}

\begin{figure}[tb]
\centering
\includegraphics[width = \columnwidth]{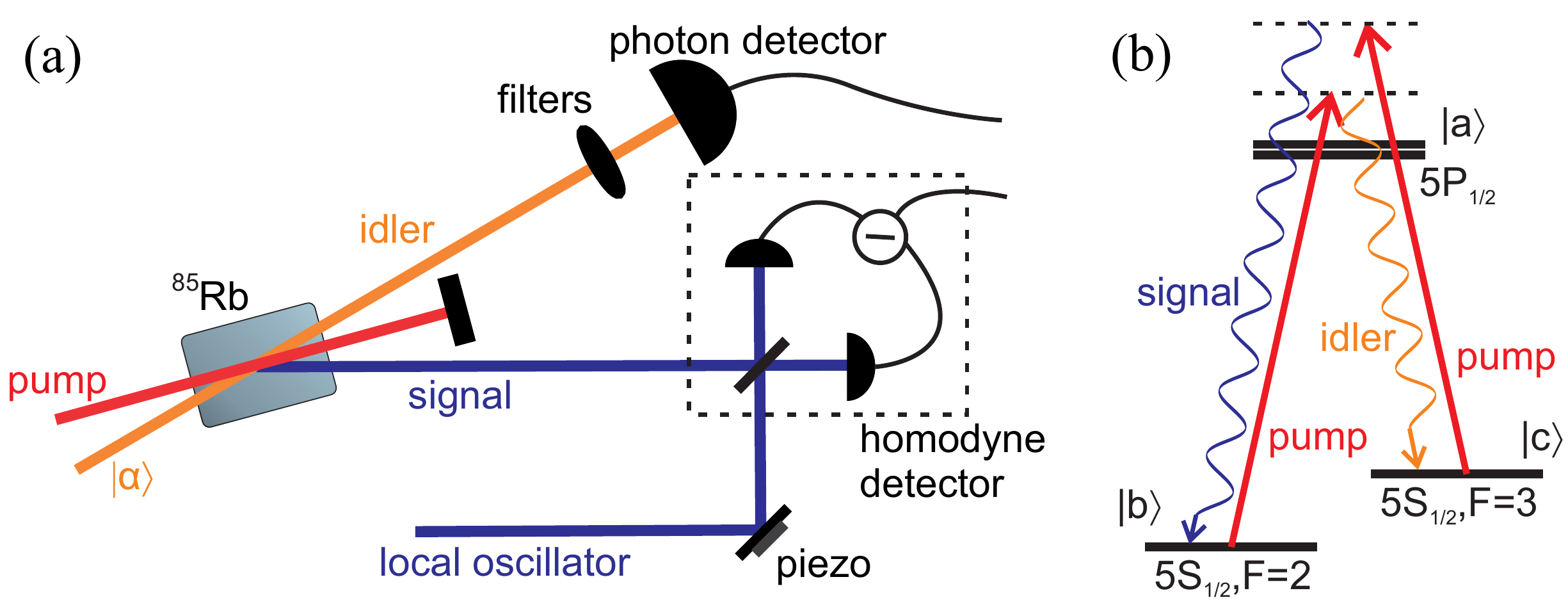}
\caption{(a) Experimental setup. A weak seed beam $|\alpha\rangle$ is overlapped along the same mode as the four-wave mixing idler and then subsequently detected after filtering. Due to the identical spatial mode and extreme time resolution of the single photon counting module, the source of the detection events is fundamentally indistinguishable, which projects the signal onto a controllable superposition state. (b) The three-level lambda system in $^{85}$Rb. A strong pump drives both Raman transitions between the $5S_{1/2},F=2$ and $F=3$ states, resulting in emission of photon pairs into the signal and idler channels.
}
\end{figure}

The experimental setup is shown in Fig.~1(a). The 1-Watt pump beam at 795 nm is generated by a Tekhnoscan TIS-SF 777 Ti:Sapphire laser and passes through a $^{85}$Rb gas cell which is heated to 107$^\circ$C. In the absence of the coherent seed beam, a quantum four-wave mixing process leads to the generation of correlated signal and idler photons along phase-matched directions \cite{lett-multimode}. A weak coherent beam $|\alpha\rangle$ is generated at the same frequency as the idler photons by double-passing a small part of the master laser field through a 1.54 GHz acousto-optical modulator. This beam is attenuated, using a series of neutral-density filters as well as waveplates and polarizing beam splitters, to the single-photon level. It is then passed through the cell in a spatial mode consistent with that of the idler photons from the four-wave mixing. Subsequently, this mode is spectrally filtered using a 55 MHz linewidth monolithic filter cavity \cite{Pan} and spatially filtered using a single mode fiber before being measured with a single photon detector.

Heralded by a click in the photon detector, we measure the state of the signal using a 100 MHz bandwidth homodyne detector \cite{HD-det}. The local oscillator is provided by a 20 mW diode laser phase-stabilized with respect to the pump using an optical phase-lock loop \cite{jurgen-pll}. The homodyne detector photocurrent is integrated over a temporal mode that is determined from the signal variance as a function of time \cite{MacRae}, to give a single quadrature value for each click event.

Remarkably, this technique automatically ensures indistinguishability between photons from the coherent state and the atomic source. This indistinguishability is not inherent: while the Raman scattering is broadband, the bandwidth of the coherent state is determined by that of the master laser, i.e. is on the order of a few kHz. However, precision timing of the photon detection events (on a scale of hundreds of picoseconds) combined with spectral filtering with a width of 55 MHz projects all photons onto indistinguishable transform limited wavepackets with the spectrum determined by the transmissivity of the spectral filter and centered in time around the detection event \cite{Aichele02}.

%Any mismatch between the bandwidths of $|\alpha\rangle$ and the photons from the four-wave mixing process is eliminated by the high timing resolution of the SPCM, which results in quantum erasure of any spectral distinguishability \cite{quantum_eraser1, quantum_eraser2}. %*Give numbers?*
%The power of $|\alpha\rangle$ is adjusted and monitored using the SPCM count rate.

%One challenge of this experiment is the requirement of eliminating the strong pump field from each of the signal and idler channels. To this end we employ a number of filtering techniques. We define the spatial modes of the idler and signal at an angle of 5 mrad from the pump beam to isolate them spatially. The idler and signal beams are orthogonally polarized with respect to the pump, allowing for polarization filtering. Furthermore, the Fabry-P\'{e}rot filter cavity in the idler channel helps to filter out any scattered pump light spectrally. The signal channel is protected from the influence of the pump due to inherent spatial and spectral filtering associated with homodyne detection.

\begin{figure}[tb]
\centering
\includegraphics[width = 0.95\columnwidth]{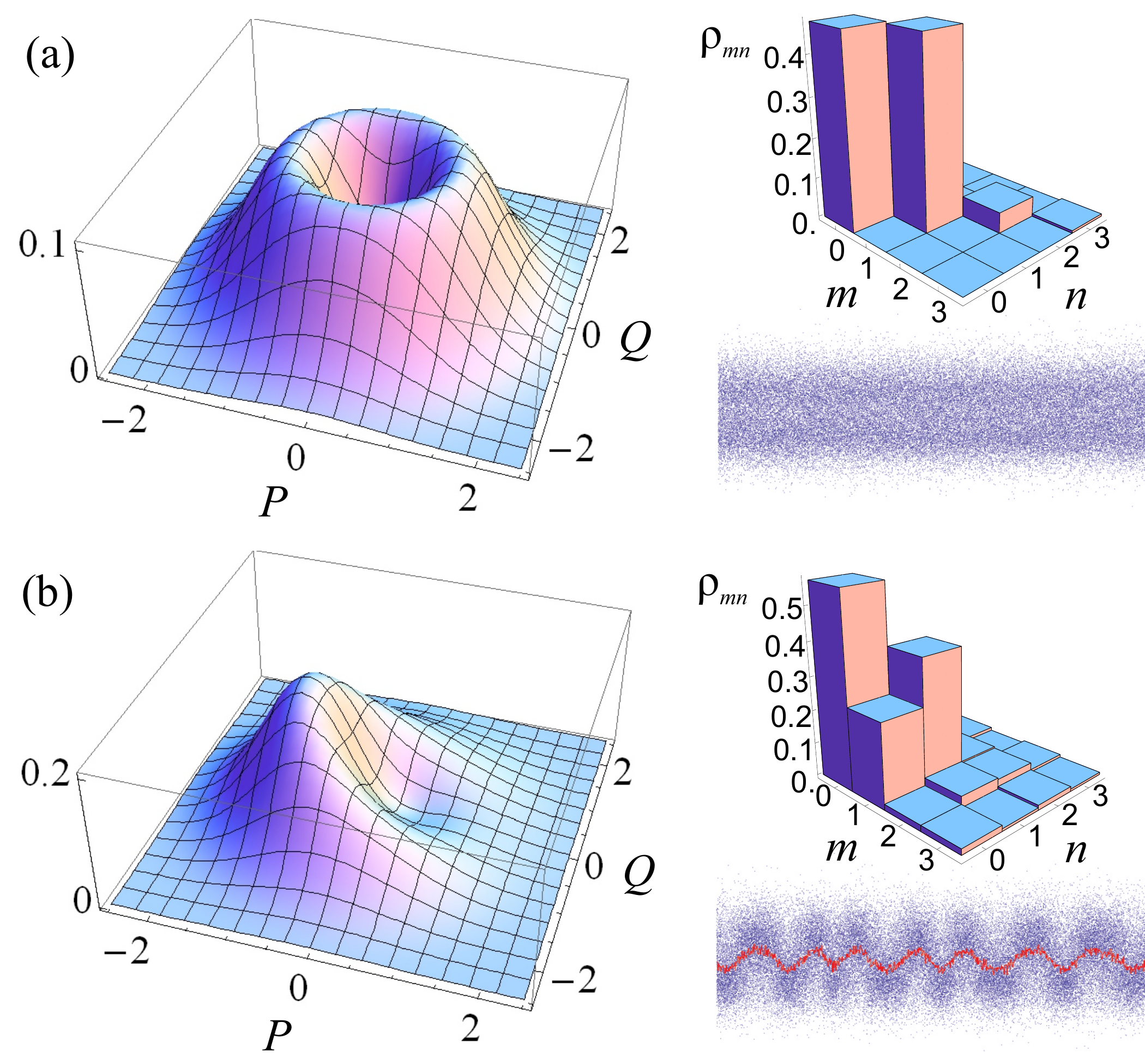}
\caption{Reconstructed quantum states. (a) The single photon Fock state obtained in the absence of the seed ($\alpha=0$). The density matrix has $\rho_{11}=0.47$, with a corresponding dip in the Wigner function at the origin and no phase dependence in the quadrature data. (b) Reconstructed state for a generated qubit, in the case where 24\% of the photon detection events are coming from $|\alpha\rangle$. Off-diagonal elements of the density matrix and phase dependence in the quadrature data indicate coherence, leading to a displacement of the peak of the Wigner function from the origin. Despite the significantly increased vacuum component, the off-diagonal terms contribute to a generalized efficiency of 46\%.} %Data: F4,Q17(Jan24)
\end{figure}

%\section{Analysis}
A piezoelectric transducer in the local oscillator path permits phase variation as required for homodyne tomography. After 100,000 quadrature values are collected, the quantum state of the signal is reconstructed using an iterative maximum-likelihood algorithm \cite{Maxlik1,Maxlik2}.

We reconstruct the density matrix of the signal state that is generated for a range of coherent state amplitudes $|\alpha|$. Fig.~2 shows a comparison between a single photon Fock state ($\alpha=0$) and a sample qubit ($\alpha\hbar/\gamma t = 0.56$). In the latter case, the off-diagonal element $\rho_{01}$ of the density matrix arises, demonstrating that the two components of the qubit are in a coherent superposition.

\begin{figure}[htb]
\centering
\includegraphics[width = \columnwidth]{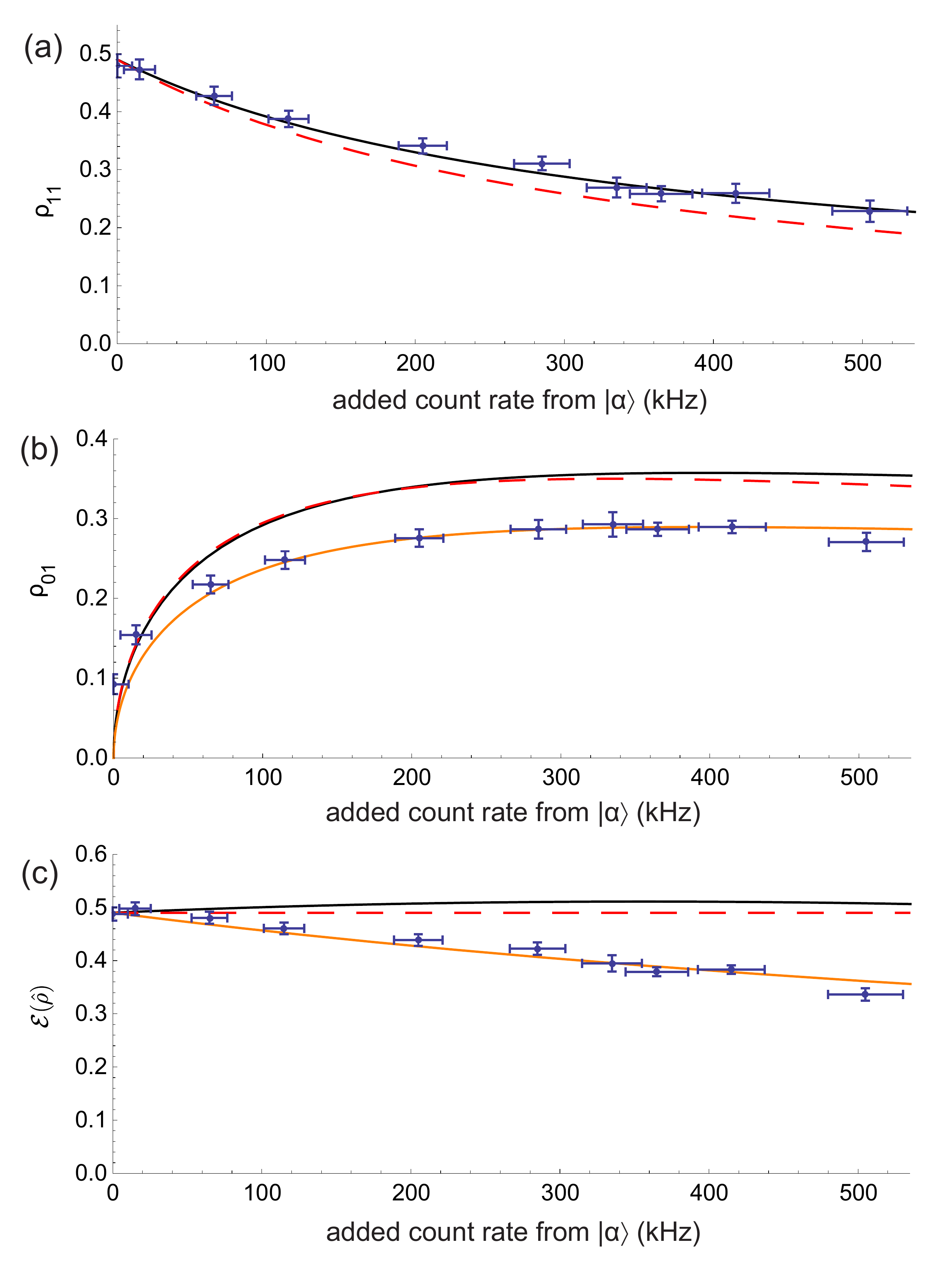}
\caption{Experimental results. Density matrix elements (a) $\rho_{11}$ and (b) $\rho_{01}$ are shown, each as a function of the added count rate in the idler channel corresponding to increasing intensity of the seed coherent state. (c) The generalized efficiency $\mathcal{E}(\hat{\rho})$ is calculated over the same range of $|\alpha|$. The black solid line is generated using a theoretical model of the four-wave mixing process considering photon number terms up to four, taking into account imperfect detection efficiency and losses in both the signal and idler channels. The dashed red curve uses the simplified model given by Eq.~(\ref{psi}) after identical losses, which neglects photon number elements above $n=1$. The orange curve considers a reduction of $\rho_{01}$ by a constant factor of 0.81 with respect to the black curve.}
\end{figure}

Fig.~3(a,b) plots elements $\rho_{11}$ and $\rho_{01}$ of the reconstructed density matrix as a function of the added count rate in the idler channel due to the seed beam. The added count rate is comparable to the count rate of 335 kHz observed at $\alpha = 0$. %As we increase $\alpha$, the off-diagonal elements of the density matrix emerge while $\rho_{11}$ decreases [Fig.3(a,b)]
The blue circles show experimentally measured data points, with error bars given by the standard deviation from multiple measurements. The dashed red curves show theoretical predictions obtained given by Eqs.~(\ref{prcount}) and (\ref{psi}), which are then subjected to linear losses along the signal channel to match experimental conditions. The solid black curves follow a similar model which takes into account additional considerations such as higher photon number components and imperfect detector efficiency in the idler channel. Limited homodyne detector bandwidth plays a significant role in the signal channel losses, as it corresponds to a timing jitter in measurements of the signal qubit. This causes a mismatch between the measured temporal mode and the actual temporal mode, leading to reduced visibility which has the same effect as spatial mode mismatch and optical losses \cite{Aichele02,Hoffman}. All of the sources of loss contribute to a combined signal channel loss of 51\%. Other best fit parameters include a combined loss of 90\% in the idler channel and a fitting parameter $\gamma$ which corresponds to $\gamma t/\hbar = 0.22$. These parameters are consistent with losses that are measured in the lab, and $\gamma$ is consistent with the magnitude of the two-photon component of the generated state \cite{MacRae}.

The quality of the generated qubits can be estimated from the experimental data by using the generalized efficiency \cite{Berry_2006,Berry_2010}, defined as the lowest possible value of $T$ such that the state observed experimentally can be obtained from another state by transmitting through a loss channel with transmissivity $T$. Neglecting the photon number terms above 1, the generalized efficiency is given by
\begin{gather} \label{eq_geneff}
\mathcal{E}(\hat{\rho}) = \frac{\rho_{11}}{1-|\rho_{01}|^2/\rho_{11}}.
\end{gather}
This quantity is displayed in Fig.~3(c). Ideally, the generalized efficiency is expected to be independent of $\alpha$. However, imperfect photon detectors and the presence of higher photon number components introduce a small dependence on $\alpha$ in the generalized efficiency.

%explanation for discrepancy
From Fig.~3, the theoretical model (black) predicts the experimental behavior for $\rho_{11}$ very well, however the experimental data for $\rho_{01}$ is below the model by what appears to be a constant factor. This indicates there is some decoherence between the $|0\rangle$ and $|1\rangle$ components of the qubit. The orange plots in Fig.~3 show the fit with $\rho_{01}$ decreased by a factor of 0.81. The source of this decoherence could not be determined, however thermal background contamination was measured to be too weak to cause this effect. This decoherence could come from some residual distinguishability between the seed beam photons and photons from four-wave mixing events, or uncertainties when reconstructing the qubit phase for each measurement event.

%collective enhancement

%The theoretical density matrix elements $\rho_{11}$ and $\rho_{01}$ after losses are then calculated accordingly. The dashed red curve shows the same model with the same loss parameter, only changing $\gamma$ to be very small $(\gamma_{red} = 0.01)$ so that the higher photon-number terms have a negligible effect.

%\section{Conclusion}
In summary, we have shown experimental creation and measurement of an arbitrary Fock-state qubit using four-wave mixing seeded by a weak coherent state. This qubit exists transiently as a CSE, and is then immediately read out optically. This scheme can be advanced from a transient CSE to demonstration of a long-lived CSE with delayed, on-demand readout, akin to a recent experiment \cite{Bimbard2014}. It can also be extended to higher photon numbers with the ultimate goal of engineering arbitrary atomic and optical states. Progress along these lines would have significant applicability for quantum light sources, quantum memories and quantum repeaters.

The authors thank Christoph Simon and Jietai Jing for their helpful suggestions. The project is supported by NSERC and CIFAR. AL is a CIFAR Fellow. ZQ is supported by the China Scholarship Council.


\begin{thebibliography}{99}
%% Do not include separate BibTeX files; if BibTeX is used,
%% paste the output (contents of .bbl file) here.

\bibitem{Lvovsky_review} A. I. Lvovsky and M. Raymer, Rev. Mod. Phys. \textbf{81,} 299 (2009).
\bibitem{Blatt13} R. Blatt, G. J. Milburn and A. I. Lvovsky, J. Phys. B: At. Mol. Opt. Phys. \textbf{46,} 100201 (2013).
\bibitem{microwave} M. Hofheinz, H. Wang, M. Ansmann, R. Bialczak, E. Lucero, M. Neeley, A. O'Connell, D. Sank, J. Wenner, J. Martinis, and A. Cleland, Nature \textbf{459,} 546 (2009).
\bibitem{trapped_ion} D. Leibfried, E. Knill, S. Seidelin, J. Britton, R. B. Blakestad, J. Chiaverini, D. B. Hume, W. M. Itano, J. D. Jost, C. Langer, R. Ozeri, R. Reichle, and D. J. Wineland, Nature \textbf{438,} 639 (2005).
\bibitem{Hammerer_review} K. Hammerer, A. S. S\o rensen and E. S. Polzik, Rev. Mod. Phys. \textbf{82,} 1041 (2010).
\bibitem{lvov_memory} A. I. Lvovsky, W. Tittel, and B. C. Sanders, Nature Photon. \textbf{3,} 706 (2009).
\bibitem{DLCZ} L. M. Duan, M. D. Lukin, J. I. Cirac, and P. Zoller, Nature (London) \textbf{414,} 413 (2001).
\bibitem{logic_gates} L. You and M. S. Chapman, Phys. Rev. A \textbf{62,} 052302 (2000).
\bibitem{metrology} S.-W. Chiow, T. Kovachy, H.-C. Chien, and M. A. Kasevich, Phys. Rev. Lett. \textbf{107,} 130403 (2011).
%\bibitem{sq1} J. Appel, P. J. Windpassinger, D. Oblak, U. B. Hoff, N. Kjaergaard, and E. S. Polzik, Proc. Natl. Acad. Sci. U.S.A. {\bf 106}, 10960 (2009).
%\bibitem{sq2} I. D. Leroux, M. H. Schleier-Smith, and V. Vuleti\'{c}, Phys. Rev. Lett. {\bf 104}, 073602 (2010).
\bibitem{sq3} T. Fernholz, H. Krauter, K. Jensen, J. F. Sherson, A. S. S\o rensen, and E. S. Polzik, Phys. Rev. Lett. \textbf{101,} 073601 (2008).
\bibitem{MacRae} A. MacRae, T. Brannan, R. Achal, and A. I. Lvovsky, Phys. Rev. Lett. \textbf{109,} 033601 (2012).
\bibitem{Polzik} S. L. Christensen, J. B. B\'{e}guin, H. L. S\o rensen, E. Bookjans, D. Oblak, J. H. M\"{u}ller, J. Appel and E. S. Polzik, New J. Phys. \textbf{15,} 015002 (2013).
%\bibitem{Kuzmich} L. Li, Y. O. Dudin and A. Kuzmich, Nature {\bf 498}, 466 (2013).
\bibitem{Bimbard} E. Bimbard, N. Jain, A. MacRae, and A. I. Lvovsky, Nature Photon. \textbf{4,} 243 (2010).
\bibitem{lett-multimode} V. Boyer, A. Marino, R. Pooser, and P. Lett, Science \textbf{321,} 544 (2008).
\bibitem{Pan} P. Palittapongarnpim, A. MacRae, and A. I. Lvovsky, Rev. Sci. Instrum. \textbf{83,} 066101 (2012).
%\bibitem{quantum_eraser1} K. De Greve, L. Yu, P. L. McMahon, J. S. Pelc, C. M. Natarajan, N. Y. Kim, E. Abe, S. Maier, C. Schneider, M. Kamp, S. H\"{o}fling, R. H. Hadfield, A. Forchel, M. M. Fejer, and Y. Yamamoto, Nature (London) {\bf 491}, 421 (2012).
%\bibitem{quantum_eraser2} M. O. Scully and K. Dr\"{u}hl, Phys. Rev. A {\bf 25}, 2208 (1982).
\bibitem{HD-det} R. Kumar, E. Barrios, A. MacRae, E. Cairns, E. H. Huntington, and A. I. Lvovsky, Opt. Commun. \textbf{285,} 5259 (2012).
\bibitem{jurgen-pll} J. Appel, A. MacRae, and A. I. Lvovsky, Meas. Sci. Technol. \textbf{20,} 055302 (2009).
\bibitem{Aichele02} T. Aichele, A. I. Lvovsky and S. Schiller, Eur. Phys. J. D \textbf{18,} 237 (2002).
\bibitem{Maxlik1} A. I. Lvovsky, J. Opt. B: Quantum Semiclass. Opt. \textbf{6,} S556 (2004).
\bibitem{Maxlik2} J. \v{R}eh\'{a}\v{c}ek, Z. Hradil, E. Knill, and A. I. Lvovsky, Phys. Rev. A \textbf{75,} 042108 (2007).
\bibitem{Hoffman} J. Appel, D. Hoffman, E. Figueroa and A. I. Lvovsky, Phys. Rev. A \textbf{75,} 035802 (2007).
\bibitem{Berry_2006} D. W. Berry, A. I. Lvovsky, and B. C. Sanders, Opt. Lett. \textbf{31,} 107 (2006).
\bibitem{Berry_2010} D. W. Berry and A. I. Lvovsky, Phys. Rev. Lett. \textbf{105,} 203601 (2010).
\bibitem{Bimbard2014} E. Bimbard, R. Boddeda, N. Vitrant, A. Grankin, V. Parigi, J. Stanojevic, A. Ourjoumtsev, and P. Grangier, Phys. Rev. Lett. \textbf{112,} 033601 (2014).

\end{thebibliography}
\end{document}